\documentclass[10pt,a4paper,twocolumn,superscriptaddress]{revtex4}

\usepackage{textcomp}
\usepackage{color}
\usepackage{amsmath}
\usepackage{mathrsfs}
\usepackage{graphicx}
\usepackage{subfig}
\usepackage{epsfig}
\usepackage{physics}
\usepackage{amssymb}
\usepackage{hyperref}
\usepackage{todonotes}
\usepackage[final]{pdfpages}

\usepackage[squaren,Gray]{SIunits}

\usepackage{pgfplots}
\pgfplotsset{compat=newest}
\usepgfplotslibrary{groupplots}
\usepgfplotslibrary{polar}
\usepgfplotslibrary{smithchart}
\usepgfplotslibrary{statistics}
\usepgfplotslibrary{dateplot}
\usepgfplotslibrary{ternary}

\begin{document}

\title{Topological Soliton Frequency Comb in Nanophotonic Lithium Niobate}

\author{Nicolas Englebert\(^{\star}\)}
\email{englebert@caltech.edu}
\affiliation{Department of Electrical Engineering, California Institute of Technology, Pasadena, California 91125, USA}

\author{Robert M. Gray\(^{\star}\)}
\affiliation{Department of Electrical Engineering, California Institute of Technology, Pasadena, California 91125, USA}

\author{Luis Ledezma}
\affiliation{Department of Electrical Engineering, California Institute of Technology, Pasadena, California 91125, USA}

\author{Ryoto Sekine}
\affiliation{Department of Electrical Engineering, California Institute of Technology, Pasadena, California 91125, USA}

\author{Thomas Zacharias}
\affiliation{Department of Electrical Engineering, California Institute of Technology, Pasadena, California 91125, USA}

\author{Rithvik Ramesh}
\affiliation{Department of Electrical Engineering, California Institute of Technology, Pasadena, California 91125, USA}

\author{Benjamin K. Gutierrez}
\affiliation{Department of Applied Physics, California Institute of Technology, Pasadena, California 91125, USA}

\author{Pedro Parra-Rivas}
\email{pedro.parra-rivas@ual.es}
\affiliation{Applied Physics, Department of Chemistry and Physics, University of Almeria, 04120 Almeria, Spain}

\author{Alireza Marandi}
\email{marandi@caltech.edu}
\affiliation{Department of Electrical Engineering, California Institute of Technology, Pasadena, California 91125, USA}
\affiliation{Department of Applied Physics, California Institute of Technology, Pasadena, California 91125, USA}

\begin{abstract} %
Frequency combs have revolutionized metrology, ranging, and optical clocks\,\cite{hall_nobel_2006,fortier_20_2019}, which have motivated substantial efforts on the development of chip-scale comb sources\,\cite{pasquazi_micro-combs_2018,gaeta_photonic-chip-based_2019}. 
The on-chip comb sources are currently based on electro-optic modulation\,\cite{yu_integrated_2022,guo_ultrafast_2023,stokowski_integrated_2024}, mode-locked lasers\,\cite{davenport_integrated_2018,hermans_high-pulse-energy_2021,guo_ultrafast_2023}, quantum cascade lasers\,\cite{hugi_mid-infrared_2012,meng_dissipative_2022,opacak_nozakibekki_2024,kazakov_driven_2025}, or soliton formation via Kerr nonlinearity\,\cite{wabnitz_suppression_1993,leo_temporal_2010,herr_temporal_2014}. However, the widespread deployment of on-chip comb sources has remained elusive as they still require RF sources, high-Q resonators, or complex stabilization schemes while facing efficiency challenges. 
Here, we demonstrate an on-chip source of frequency comb based on the integration of a lithium niobate nanophotonic circuit with a semiconductor laser that can alleviate these challenges.
For the first time, we show the formation of temporal topological solitons in a on-chip nanophotonic parametric oscillator with quadratic nonlinearity and low finesse. These solitons, independent of the dispersion regime, consist of phase defects separating two $\pi$-out-of-phase continuous wave solutions at the signal frequency, which is at half the input pump frequency\,\cite{trillo_stable_1997,parra-rivas_frequency_2019}. 
We use on-chip cross-correlation for temporal measurements and confirm formation of topological solitons as short as 60\,fs around 2\,$\mu$m, in agreement with a generalized parametrically forced Ginzburg-Landau theory\,\cite{leo_walk-off-induced_2016,mosca_modulation_2018,parra-rivas_dissipative_2022}.
Moreover, we demonstrate a proof-of-concept turn-key operation of a hybrid-integrated source of topological frequency comb.
Topological solitons offer a new paradigm for integrated comb sources, which are dispersion-sign agnostic and do not require high-Q resonators or high-speed modulators and can provide access to hard-to-access spectral regions, including mid-infrared\,\cite{schliesser_mid-infrared_2012}.
\end{abstract}

\maketitle
\def\thefootnote{$\star$}\footnotetext{These authors contributed equally to this work}\def\thefootnote{\arabic{footnote}}

\begin{figure*}  
    \centering
    \hspace{-5mm}
    \includegraphics[width=\linewidth]{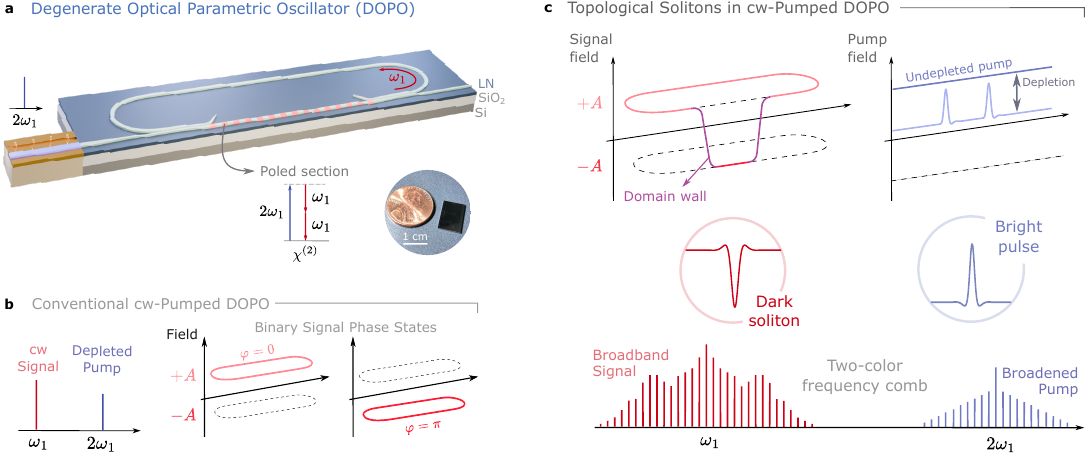}
    \caption{\textbf{Topological soliton concept.} \textbf{a},\,Schematic of the hybrid-integrated topological soliton frequency comb source. A degenerate optical parametric oscillator (DOPO) is pumped by a continuous wave (cw) laser diode. The poled section is phase-matched for degenerate parametric down-conversion, leading to the generation of signal photons at half the pump frequency. Inset: image of the actual chip containing 30 OPOs.     
    \textbf{b},\,The cw pumping of a DOPO results in a cw signal in the conventional regime. Owing to a $\pi$ phase indetermination, the signal field sign is either positive ($+A_h^+$) or negative ($-A_h^+$), with equal probability. 
    \textbf{c},\,These two fields with opposite signs can coexist through the formation of domain walls connecting the opposite signs. The intensity profiles of these domain walls correspond to dark pulses named topological solitons. At the pump, they coincide with the formation of bright pulse. This dark-bright pulse pair forms a two-color frequency comb.}
    \label{fig:Panel1}
\end{figure*}

Optical frequency combs have revolutionized many fields of science\,\cite{fortier_20_2019}.
A common type of on-chip sources of frequency combs have been ultra low-loss (high Q) driven optical cavities with a third-order (Kerr, $\chi^{(3)}$) nonlinearity. 
Through the formation of Kerr cavity solitons\,\cite{wabnitz_suppression_1993,leo_temporal_2010,herr_temporal_2014}, they can lead to octave-spanning combs around the driving frequency. 
Integrated electro-optic (EO) combs have recently appeared as an alternative on-chip comb source, benefiting from the recent advances in nanophotonic circuits on thin-film materials with quadratic nonlinearity ($\chi^{(2)}$). Leveraging the Pockels effect, EO combs offer an efficient way to generate broadband frequency combs through sideband generation, using radio-frequency input\,\cite{zhang_broadband_2019,yu_integrated_2022,zhang_ultrabroadband_2025}. 
Using such an EO modulator inside a laser,\cite{guo_ultrafast_2023} or a quadratic resonator,\cite{stokowski_integrated_2024} has also been studied as another variant of on-chip comb sources, alongside quantum cascade lasers, leveraging a large effective Kerr effect to generate mid-infrared frequency combs\,\cite{meng_dissipative_2022,opacak_nozakibekki_2024,kazakov_driven_2025}. 

While so far Kerr nonlinearity have been the dominant mechanism for soliton-based comb sources, resonators with strong quadratic nonlinearity can host a wide range of fundamentally different types of solitons\,\cite{buryak_optical_2002,parra-rivas_frequency_2019,villois_soliton_2019,nie_photonic_2022,skryabin_coupled-mode_2020,parra-rivas_dissipative_2022}. Such \textit{quadratic solitons}  offer advantages for comb sources: they can form in low-Q resonators irrespective of the chromatic dispersion regime, and can access spectral regions beyond the reach of standard laser sources without reliance on fast modulators.
Despite their potential and some recent demonstrations in bulk\,\cite{jankowski_temporal_2018,musgrave_dissipative_2025} and hybrid resonators\,\cite{odonnell_widely_2020,roy_temporal_2022}, quadratic soliton microcombs have so far been limited to the second-harmonic enhanced configuration\,\cite{lu_two-colour_2023} or relied on the intrinsic Kerr effect\,\cite{bruch_pockels_2020} and have not been realized as fully integrated sources. 

In this work, we demonstrate for the first time a new kind of hybrid-integrated frequency comb source based on quadratic nonlinearity in both dispersion regimes using low-finesse and phase-matched degenerate optical parametric oscillators in lithium niobate nanophotonics. 
The mode-locked comb spontaneously forms at a center wavelength of twice that of the pump and does not require contribution from the Kerr effect\textemdash intrinsic or effective\,\cite{parra-rivas_dissipative_2022} or EO modulators.
Specifically, it is sustained by dark pulses that consist of topological phase defects separating two $\pi$-out-of-phase continuous wave solutions. Such \textit{topological soliton} are robust solitary waves whose temporal profile is determined by a balance between dispersion and quadratic nonlinearity\,\cite{trillo_stable_1997,parra-rivas_frequency_2019}. 

These topological solitons are distinct from the topological behaviors studied in the real or synthetic networks of resonators, including topological optical parametric oscillators (OPOs)\,\cite{roy_topological_2022} as well as the formation of frequency combs\,\cite{flower_observation_2024} and/or Kerr solitons\,\cite{tikan_emergent_2021} in them.
They share similarities with the phase solitons found in driven non-parametric systems such as semiconductor lasers\,\,\cite{gustave_dissipative_2015}, quantum cascade lasers\,\cite{prati_soliton_2021}, and Kerr resonators where they are associated with spontaneous symmetry breaking\,\cite{garbin_dissipative_2021,coen_nonlinear_2024}. 

Spatial topological solitons have been studied in degenerate OPOs (DOPOs)\,\cite{trillo_stable_1997,oppo_domain_1999} and four-wave-mixing oscillators\,\cite{taranenko_pattern_1998,esteban-martin_control_2005}. 
However, despite substantial work on DOPOs and several theoretical studies\,\cite{parra-rivas_frequency_2019,nie_photonic_2022,sanchez_ultrashort_2023}, \textit{temporal} topological solitons have not yet been experimentally demonstrated due to the challenges involved in precisely controlling phase matching and the round-trip phase, as well as the complexity associated with their temporal characterization. 

We have broken this barrier by leveraging the nanophotonic lithium niobate platform for the generation and temporal characterization of topological solitons using an integrated optical parametric amplifier-based cross-correlation\,\cite {zacharias_energy-efficient_2025}. Moreover, we demonstrate a turn-key hybrid-integrated frequency comb source based on topological solitons, highlighting a path toward compact and scalable sources that can break the wavelength coverage of the current technologies.

\section*{Results}

Our on-chip comb source is based on the hybrid integration of a pump laser and a degenerate optical parametric oscillator (DOPOs), as depicted in Fig.\,\ref{fig:Panel1}a. An image of the actual chip, containing 30 OPOs, is given in inset.
The poled section is phase-matched for degenerate parametric down-conversion, where the pump photons at frequency $2\omega_1$ are converted into pairs of signal photons ($\omega_1$). This process is phase-sensitive and guarantees a fixed signal-pump relative phase\,\cite{byer1975optical}. 
The DOPO couplers are designed such that only the frequencies in the vicinity of the signal are coupled into the resonator while the pump remains in the straight section.
Such DOPOs have been recently realized and operated in conventional OPO operation regimes\,\cite{ledezma_octave-spanning_2023,roy_visible--mid-ir_2023,kellner_low_2025}.
Figure\,\ref{fig:Panel1}b illustrates such a regime under continuous wave (cw) pumping, where the pump is depleted in favor of a cw signal field which can take one of the binary phase states\,\cite{byer1975optical}. 
These two equiprobable solutions ($\pm A_h^+$, Fig.\,\ref{fig:Panel1}c), have already been leveraged for quantum random number generation\,\cite{marandi_all-optical_2012,gray_large-scale_2024} and photonic Ising machines\,\cite{marandi_network_2014} and are the key to generating broadband frequency combs in this work. When these two solutions of opposite signs coexist, owing to the resonator's periodic boundary conditions, \textit{domain walls} (DWs) form in the signal wave.
As illustrated in Fig.\,\ref{fig:Panel1}c, the formation of such domain walls corresponds to intensity profiles of dark pulses coined topological solitons (TS)\,\cite{trillo_stable_1997}. 
Due to local non-depletion, TSs coincide with the formation of bright pulses at the pump frequency, resulting in a two-color frequency comb.
\begin{figure*}  %%% Fig2
    \centering
   % \vspace{2mm}
    \hspace{-5mm}
    \includegraphics[width=\linewidth]{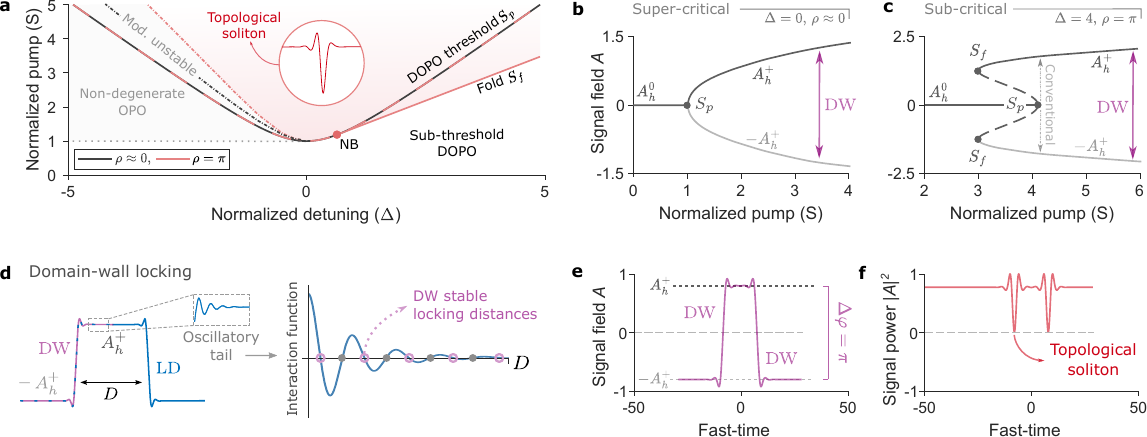}
    \caption{\textbf{Theoretical analysis of temporal topological solitons}. 
    \textbf{a}, Phase diagram showing the main bifurcation lines of the system for two values of phase mismatch [$\varrho\approx0$ (gray) and $\varrho=\pi$ (red)], $\eta_1=-1$ and $\eta_2=0.13$. $S_P$: pitchfork bifurcation.  $S_f$: fold. The pink (white) area indicates the existence of stable topological solitons (trivial solution). Gray area: modulationally unstable regions, delimited by point-dashed lines (see main text). 
    \textbf{b}, Bifurcation diagram in the supercritical configuration for $\Delta=0$ and $\varrho\approx0$. The purple arrow suggests the formation of domain walls (DWs). Solid (dashed) lines represent stable (unstable) states. We omit the (unstable) trivial solution above the threshold. \textbf{c}, Similar to (b) but for the subcritical regime with $\Delta=4$ and $\varrho=\pi$. Gray arrows: close to $S_p$, the system tends to dynamically converge toward the conventional regime ($A_h^+$ \textit{or} $-A_h^+$) rather than forming DWs ($A_h^+$ \textit{and} $-A_h^+$).
    \textbf{d}, Schematic representation of a localized domain arising from the locking between two DWs through their oscillatory tails. The interaction function governs the locking distances (see SI, section IV).
    \textbf{e},\,Field envelope and intensity profile (\textbf{f}) of a TS pair in the supercritical regime [see (b)] for $S=2$.% \textbf{f}, Corresponding intensity profile.
    }
    \label{fig:Panel2}
\end{figure*}

\begin{figure*}  
    \centering
    \hspace{-5mm}
    \includegraphics[width=0.9\linewidth]{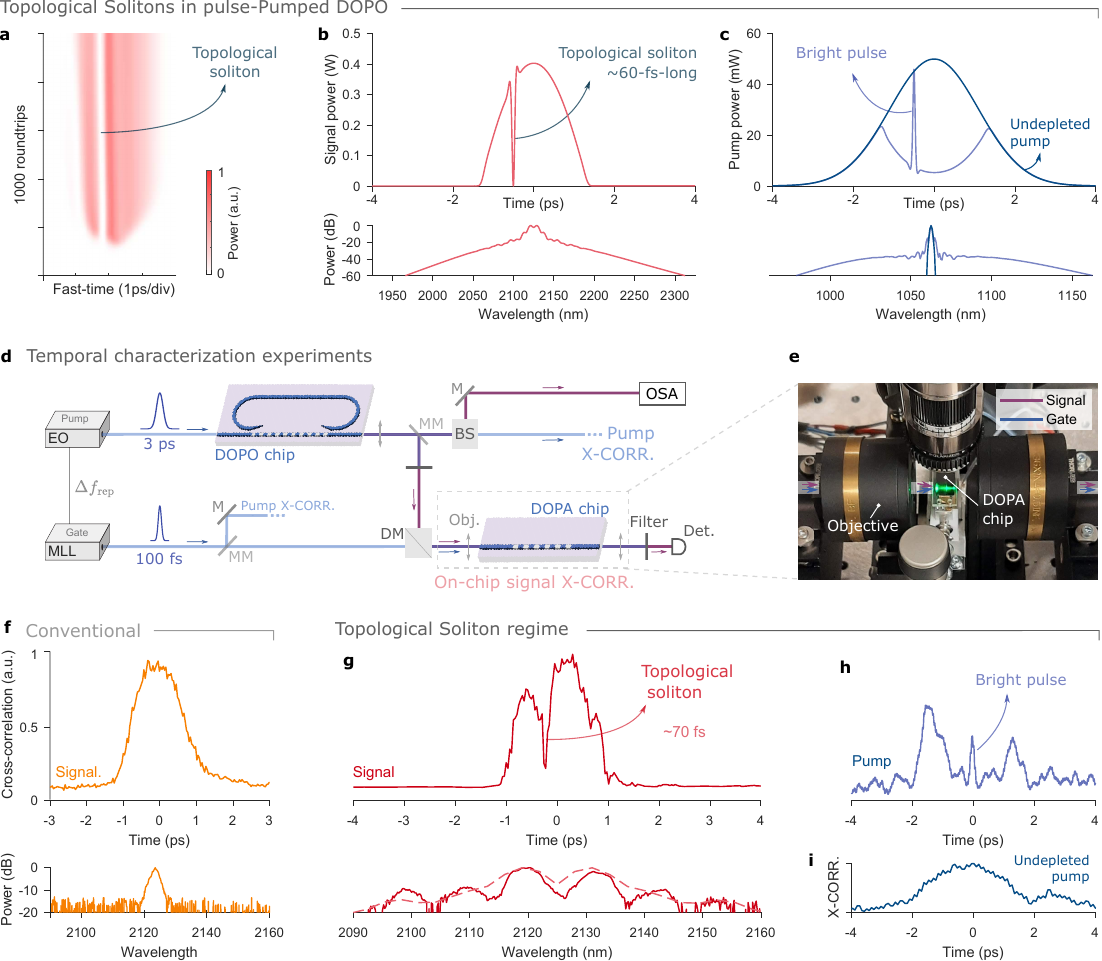}
    \caption{\textbf{Temporal and spectral characteristics of topological solitons}. 
    \textbf{a},\,Pulsed pumping allows the spontaneous formation of an isolated stable topological soliton (TS) from quantum noise ($S=1.6$, $\Delta = 0$, $\rho \approx 0$, $\Delta t = 95\,$fs/roundtrip). 
    \textbf{b},\,Theoretical temporal profile of the signal (top) and spectrum (bottom) after 100,000 roundtrips. 
    \textbf{c},\,Temporal profile of the pump (light blue) at the same condition as (b). Dark blue: undepleted pump profile. 
    \textbf{d},\,\,Schematic of the experimental setup (see Methods). M: mirror; MM: magnetic mirror; BS: beamsplitter; DM; dichroic mirror. DOPA: degenerate optical parametric amplifier. 
    \textbf{e},\,Photograph of the DOPA chip. The green emission corresponds to the second harmonic generation of the gate pulses on the chip.
    \textbf{f},\,Experimental cross-correlation (top) and spectrum (bottom) when the DOPO is synchronously pumped ($\Delta t = 0$) near resonance and perfect phase-matching. 
    \textbf{g},\,By detuning the pump from the synchronous condition ($\Delta t \approx 95\,$fs/roundtrip), a 70-fs-long dark pulse is formed in the output signal. Dashed: theoretical spectrum [see (b)]. The dark pulse in the signal coincides with the formation of a bright pulse at the pump (\textbf{h}), resulting from a depletion above 50\% (\textbf{i}).}
    \label{fig:Panel3}
\end{figure*}

\vspace{-3mm}
\subsection*{Non-local parametrically forced Ginzburg-Landau theory}
\begin{figure*} 
    \centering
    \hspace{-5mm}
    \includegraphics[width=\linewidth]{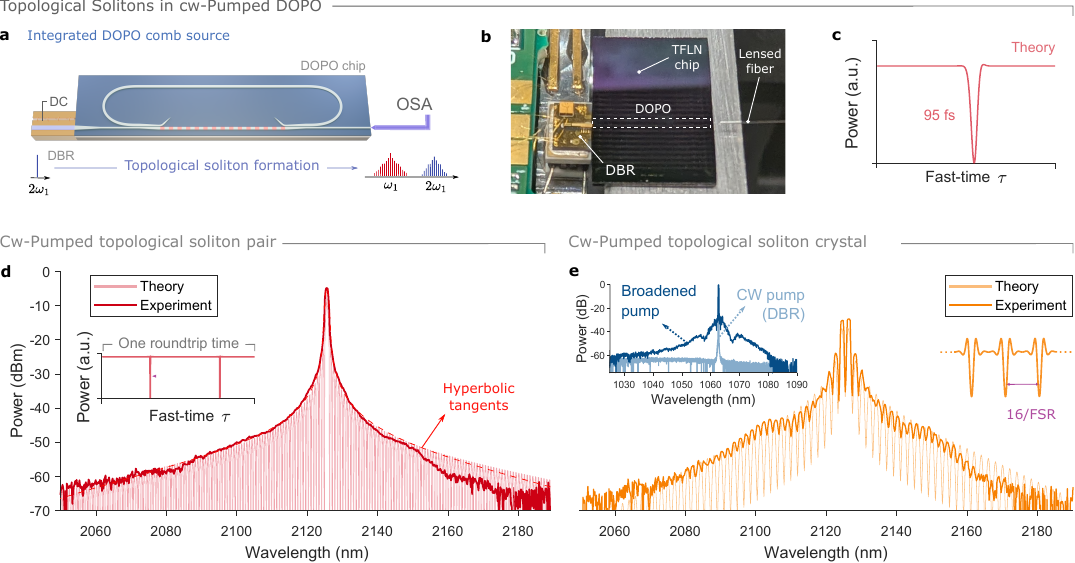}
    \caption{\textbf{Integrated topological soliton frequency comb source}. 
    \textbf{a},\,Experimental setup for the formation and characterization of the topological soliton frequency comb.
    \textbf{b},\,An electrically-driven DBR singly-frequency laser is edge-coupled to a DOPO nanophotonic chip.
    \textbf{c},\,Under cw-pumping, the numerical integration of Eq.\,\eqref{eq:GLE} reveals the spontaneous formation of 95-fs-long topological soliton using the experimental parameters ($S=1.1$, $\Delta \approx 0$, $\rho\approx0$). 
    \textbf{d},\,Experimental topological soliton frequency comb (red) and theory [light red, see (c)]. Dashed: theoretical spectrum resulting from a pair of 95-fs-long hyperbolic tangents. 
    \textbf{e},\,Experimental multi-TS state (orange). Light orange: simulation of a perfect TS crystal with a dark pulse spacing of $D\approx7.75\,$ps. Inset: DBR spectrum and its broadening in the topological soliton regime.
    }
    \label{fig:Panel4}
\end{figure*}

The formation of TSs has been theoretically studied in spatial\,\cite{trillo_stable_1997,oppo_domain_1999} and pump-resonant DOPOs\,\cite{parra-rivas_frequency_2019}. Here, we extend the theory to the temporal case where only the signal resonates using a non-local parametrically forced Ginzburg-Landau equation\,\cite{leo_walk-off-induced_2016,mosca_modulation_2018,parra-rivas_dissipative_2022} [see Supplementary Information (SI), section I]:
\begin{equation}
    \partial_t \mathsf{A} = -(1+i\Delta)\mathsf{A} - i\eta_1\partial_\tau^2\mathsf{A} - \mathsf{A}^*(\mathsf{A}^2\otimes I) + S\mathsf{A}^*,
    \label{eq:GLE}
\end{equation}
where $t$ is a slow time describing the evolution of the intracavity normalized signal envelope $\mathsf{A} = A_{\omega_1}(t,\tau)$ over consecutive roundtrips with $\tau$, a fast time defined in a reference frame traveling at the signal group velocity. $\Delta$ is the normalized cavity phase detuning, $\eta_1$ indicates the dispersion regime (+1: normal, -1: anomalous), $S$ is the normalized pump field amplitude, and $\otimes$ denotes the convolution with the nonlocal kernel $I$\,\cite{leo_walk-off-induced_2016,mosca_modulation_2018,parra-rivas_dissipative_2022}. The kernel depends on the phase mismatch ($\varrho$), the group velocity mismatch (GVM) between the pump and the signal, and the pump depletion and group velocity dispersion (GVD).  In what follows, we take $\eta_1=-1$ and $\eta_2=0.13$, the normalized pump-to-signal GVD ratio (see SI, section I), to match the experimental values reported in Figs.\,\ref{fig:Panel3} and \ref{fig:Panel4}. 

Topological solitons arise from the connection between two $\pi$-out-of-phase stable cw solutions. Equation \eqref{eq:GLE} admits non-trivial cw solutions $\mathsf{A}^\emptyset_h$ that emerge from a Pitchfork bifurcation at the trivial state ($\mathsf{A}_h^0=0$) at the oscillation threshold, $S_p\equiv \sqrt{1+\Delta^2}$. This bifurcation line, shown in the $(\Delta,S)$-phase diagram in Fig.~\ref{fig:Panel2}a, can emerge in two distinct ways from the trivial state, leading to different bistability regions and TSs. On the one hand, the bifurcation is \textit{super-critical} when two non-trivial states $\pm\mathsf{A}^\emptyset_h$ emerge from the oscillation threshold $S_p$ as depicted in Fig.~\ref{fig:Panel2}b for $\Delta=0$. On the other hand, the bifurcation can be \textit{sub-critical}, as illustrated in Fig.~\ref{fig:Panel2}c for $\Delta=4$. In this case, Eq.\,\eqref{eq:GLE} admits four non-trivial states, two of which ($\pm\mathsf{A}^-_h$) bifurcate at the oscillation threshold to merge with the others ($\pm\mathsf{A}^+_h$) at a fold $S_f$. The transition between the two types of bifurcations depends on the phase-mismatch-dependent parameter $\beta(\varrho)$ and occurs at the so-called nascent bistability point (Fig.~\ref{fig:Panel2}a, red dot). To find where TSs exist, we then perform a linear stability analysis of all the cw solutions.
%and mark their stability using solid (dashed) lines for stable (unstable) solutions.
%
The pink shadowed area in the $(\Delta,S)$-plane summarizes the analysis, highlighting where topological solitons, i.e., the DWs $\mathsf{A}^+_h\leftrightarrow-\mathsf{A}^+_h$, exist and are stable.
For positive detuning, the non-trivial cw solutions are stable against small perturbations. However, for $\Delta<0$, the trivial state $\mathsf{A}_h^0$ is modulationally unstable above $S=1$ (gray dotted line), corresponding to non-degenerate emission. 
Similarly, above $S_p$, the non-trivial cw solutions are unstable up to a given value of $S$, as indicated by the point-dashed lines. More details are given in the SI (section III). 

So far, we have only considered isolated TSs. Yet, owing to the cavity boundary conditions, they must necessarily form in pairs that will interact through their tails. The tail's shape is governed by an interaction function that depends on the DOPO parameters and is plotted in Fig.\,\ref{fig:Panel2}d. When the tails monotonically decay, the two TSs attract each other until they eventually annihilate in a process known as coarsening\,\cite{allen_microscopic_1979}. This leads to the conventional operation regime of DOPOs. In contrast, oscillating tails create regions where two DWs can lock at a quantized separation distance $D$ to form a localized domain (Fig.\,\ref{fig:Panel2}e), allowing the TS pair to persist (Fig.\,\ref{fig:Panel2}f). We stress that these interactions can lead to rich nonlinear dynamics and various temporal profiles. Several bifurcation plots are provided in the SI (sections IV-V), along with additional details on the interaction function and the coarsening regions. 

To verify these predictions, we numerically integrate Eq.\,\eqref{eq:GLE} using the parameters of Figs.\,\ref{fig:Panel2}b-c, slightly above the oscillation threshold and starting from quantum noise. While the super-critical case strongly favors the topological soliton regime, the sub-critical case, although it also supports TSs, leads to the conventional regime most of the time. In the former, $\pm A_h^+$ simultaneously emerge at the oscillation threshold $S_p$, favoring the formation of DWs. 
In contrast, in the sub-critical scenario, the system converges to one of the two cw solutions ($\pm A_h^+$), as shown by the point dash arrows in Fig.\,\ref{fig:Panel2}c. Supporting statistical data are shown in the supplementary Fig.\,S10.
Therefore, to access the TS regime and avoid 
regions of parametric instabilities and coarsening in our system, we increase the pump power while keeping the DOPO near resonance ($\Delta\approx0$) and perfect phase-matching ($\rho\approx0$).

\subsection*{Topological soliton state}
In a first series of experiments, use pulse pumping\,\cite{roy_visible--mid-ir_2023} to study the formation of isolated topological solitons, along with their temporal and spectral characterization. Numerical simulations of Eq.\,\eqref{eq:GLE} show that, even in the super-critical case (see, e.g., Fig.\,\ref{fig:Panel2}b), when the pulsed pump repetition frequency perfectly matches the cavity free spectral range, the steady-state corresponds to the conventional regime\,\cite{roy_visible--mid-ir_2023}. This is due to a small difference between the TS and the degenerate signal group velocities\,\cite{coullet_breaking_1990,gomila_theory_2015}.
%, causing the TSs to drift and eventually disappear.
%
In contrast, introducing a slight repetition rate mismatch $\Delta t$ allows the formation of a single TS that indefinitely persists, as illustrated by the DOPO simulation plotted as a 2D colormap in Fig.\,\ref{fig:Panel3}a. The steady-state DOPO signal consists of a single 60-fs-long topological soliton(Fig.\,\ref{fig:Panel3}b), which coincides with the formation of a bright pulse at the pump frequency (Fig.\,\ref{fig:Panel3}c).

To unambiguously confirm these predictions, we carry out comprehensive temporal measurements. So far, accurately resolving the temporal profiles of femtosecond dark pulses generated on chip has proven to be challenging. Several techniques have been investigated, such as spectral line-by-line pulse shaping\,\cite{ferdous_spectral_2011} or self-cross-correlation\,\cite{xue_mode-locked_2015}. The lack of broadband amplifiers at the signal frequency prevents their use here. 
To overcome these challenges and reveal the signal's temporal profile, we utilize a degenerate optical parametric amplifier (DOPA) in thin-film lithium niobate (TFLN) within a dual-comb cross-correlation setup, depicted in Fig.\,\ref{fig:Panel3}d\,\cite{zacharias_energy-efficient_2025}. A photo of the DOPA chip is shown in Fig.\,\ref{fig:Panel3}e.
The DOPO chip is also made in TFLN\,\cite{ledezma_octave-spanning_2023,roy_visible--mid-ir_2023}, has a net anomalous dispersion ($\eta_1=-1$) and roundtrip losses of 30\%, corresponding to a finesse $\mathcal{F}\approx18$ (see Methods). We control its temperature and confirm near-perfect phase-matching ($\rho\approx0$) through optical parametric generation\,\cite{ledezma_intense_2022}.

We first pump the DOPO chip with 3-ps-long pulses synchronized to the cavity free spectral range (8.158\,GHz). To ensure operation in the super-critical regime (see Fig.\,\ref{fig:Panel2}b), we scan the pump wavelength to identify the nearest degenerate resonance. We then turn the pump off, set its center frequency to verify $\Delta \approx 0$, and turn it on again. For this purpose, we use the quasi-pulsed pump scheme\,\cite{ledezma_octave-spanning_2023}. It is equivalent to repeating the same experiment continuously while ensuring the system is in its steady state more than 90\% of the time. The results are shown in Fig.\,\ref{fig:Panel3}f. As expected, both temporal and spectral measurements confirm the repeated operation in the conventional regime. 
We then repeat the same experiment, but this time with a repetition rate mismatch of $\Delta t\approx 95\,$fs/roundtrip. The results are reported in Fig.\,\ref{fig:Panel3}g. In stark contrast with the conventional regime, we observe the formation of a single 70-fs-long dark pulse, which agrees well with the theoretical predictions. 
We also measure the pump temporal profile in the topological soliton regime with a standard table-top SHG cross-correlation setup\,\cite{trebino_frequency-resolved_2000}. The pump cross-correlation plotted in Fig.\,\ref{fig:Panel3}h reveals the formation of a bright pulse, 
coinciding with the dark pulse at the signal frequency. Comparison with the undepleted pump reveals an overall pump depletion of above 50\% (Fig.\,\ref{fig:Panel3}i).
The results related to the formation of a topological soliton frequency comb in the pulse-pumped regime with a net normal dispersion ($\eta=+1$) are given in the SI (section VII).

\subsection*{Integrated topological soliton comb source}
After establishing the formation of topological solitons using pulsed pumping scheme, we study the formation of topological solitons with a cw pump to illustrate the potential of topological solitons as scalable integrated frequency comb sources.
We use a distributed Bragg reflector (DBR) single-frequency laser to directly pump the same DOPO chip via edge coupling, as shown in Figs.\,\ref{fig:Panel4}a-b. The DBR is electrically driven by a DC current source, and the DOPO output is collected with a lensed tip fiber. We use the same integrated DOPO as for the temporal characterizations.
Figure \,\ref{fig:Panel4}c shows the theoretical prediction of the TS state obtained from Eq.\,\eqref{eq:GLE} with these experimental conditions. The topological solitons are predicted to be 95-fs long. 

In experiments, we first identify the DC current that leads to the conventional regime, associated with a sub-critical bifurcation of the non-trivial cw solutions (see Fig.\,\ref{fig:Panel2}c). By slightly adjusting the current around this value and switching the DBR on and off, a broad spectrum is obtained, as shown in Fig.\,\ref{fig:Panel4}d. The measured spectrum is plotted alongside the theoretical predictions of Eq.\,\eqref{eq:GLE} for a pair of 95-fs topological solitons. The experimental spectrum, the theoretical prediction, and the simplified hyperbolic tangent shape serve as strong evidence for the operation of the DOPO in the TS regime.
By further detuning the current, we observe a comb state indicative of a TS crystal (Fig.\,\ref{fig:Panel4}e). This spectrum has two distinctive features. First, the spectral fringes suggest a TS spacing of exactly one-sixteenth of the round-trip time. Second, the degenerate frequency $\omega_1$ is attenuated by 20\,dB with respect to its two neighboring comb lines, indicating uniform distribution of TSs in the time domain. As shown in the figure, these behaviors are closely reproduced by our simulation of Eq.\,\eqref{eq:GLE}. We also observe a significant pump broadening in the topological soliton regime (Fig.\,\ref{fig:Panel4}e, inset).

\subsection*{Discussion and outlook}
Our results mark the first demonstration of temporal topological solitons in nanophotonics and how they can be utilized as an integrated source of broadband frequency combs.
%
%So far, integrated comb sources have either relied on the Kerr effect--- whether intrinsic \cite{bruch_pockels_2020} or effective \cite{lu_two-colour_2023}--- or on electro-optic modulation\,\cite{yu_integrated_2022,guo_ultrafast_2023,stokowski_integrated_2024}. Our work presents a new paradigm for hybrid-integrated quadratic frequency comb sources that neither depends on any of these effects nor requires complex stabilization schemes. Specifically, we demonstrate the existence of dark pulses--- topological solitons\,\cite{trillo_stable_1997}--- in on-chip DOPOs.
%
We show that these quadratic solitons spontaneously emerge above the oscillation threshold, regardless of the resonator Q factor and the dispersion regime (see SI section VII) , which will enable frequency comb generation in a wide range of wavelengths outside the typical near-IR window. 
While our demonstration is on thin-film lithium niobate nanophotonics, our results can be transposed to other integrated $\chi^{(2)}$ platforms, such as aluminum nitrate\,\cite{bruch_pockels_2020} or lithium tantalate\,\cite{wang_lithium_2024}, as well as free-space\,\cite{roy_temporal_2022} and fiber-based\,\cite{englebert_parametrically_2021} DOPOs.
Our results unveil the rich landscape of topological soliton formation through bifurcation analysis of a generalized parametrically forced Ginzburg-Landau equation and provide a simple physical intuition. Our theoretical results can serve as a foundation for future investigations, particularly for topological soliton crystals, sensitivity and noise analysis, as well as operation in the few-cycle regime.
Furthermore, the combination of topological solitons with readily available modulators\,\cite{wang_integrated_2018} and poled waveguides\,\cite{ledezma_intense_2022} on the same platform can benefit several applications, such as LiDAR\,\cite{riemensberger_massively_2020} and coherent telecommunications\,\cite{marin-palomo_microresonator-based_2017}. 
More advanced nanophotonic circuits, for instance involving delay lines\,\cite{marandi_network_2014} or coupled resonators\,\cite{roy_non-equilibrium_2023}, can also be envisaged in the future, either to increase the comb formation efficiency or to force the deterministic formation of a specific localized domain.

\small
\section*{Methods}
\subparagraph*{\hskip-10pt DOPOs design and fabrication}\ \\
\noindent All the results presented in the main text are obtained from a single integrated DOPO with a net anomalous dispersion ($\eta_1=-1$). The formation of a topological soliton frequency comb in the normal dispersion regime, using another chip, is shown in the supplementary Fig.\,S9. The DOPO with a net anomalous dispersion is fabricated using an 800-nm-thick MgO-doped lithium niobate layer and a SiO2 buffer layer on x-cut commercial wafers (NANOLN). Its waveguides have a top width of 1480\,$\mu$m and 455\,nm of etching depth. The waveguides are patterned by e-beam lithography and dry etched with Ar+ plasma. After etching, we add a 1\,um-thick silica cladding through plasma-enhanced chemical vapor deposition. We achieve quasi-phase matching in a $L_2=5$-mm-long region through periodic poling\,\cite{ledezma_intense_2022}. To compensate for the thin film lithium niobate thickness variations that can jeopardize phase-matching, we fabricated OPOs with poling periods ranging from 5.65\,$\mu$m to 5.95\,$\mu$m in 10-nm steps. We include a straight waveguide next to each OPO for calibration and quasi-phase matching verification\,\cite{ledezma_octave-spanning_2023}.
The DOPO has a pair of wavelength-selective couplers to prevent the pump from resonating and ensure the degenerate resonant operation. The input and output couplers are identical and designed so that the wavelengths above 1.8\,$\mu$m have large coupling factors ( $>80$\%). In contrast, those around the pump wavelength are barely coupled to the resonator. For cw-pumping operation, the top width of the DOPO input waveguide is tapered out to minimize the coupling loss between the TFLN chip and the single-frequency laser diode (Thorlabs DBR1064PN).
The DOPO has a signal GVD of $\beta_2 = -25\,$fs$^2$/mm, a pump GVD of $\beta_{2p} = 50\,$fs$^2$/mm, and a GVM between the signal and the pump of $\Delta\beta_1=3$\,fs/mm. The second-order nonlinear effective parameter is estimated to $\kappa=300\,$W$^{-0.5}$.m$^{-1}$. This yields the normalized parameters: $\eta_1 = -1$, $\eta_2 = 0.13$, and $d=0.77$, which we set to zero in the theoretical analysis.

\subparagraph*{\hskip-10pt Experimental setup}\ \\
To study the topological soliton formation in the pulsed pump regime, we pump the DOPO with 3-ps-long pump pulses, produced by an electro-optic (EO) frequency comb\,\cite{roy_visible--mid-ir_2023}. It is generated by cascading the output of a CW laser by one intensity modulator and three phase modulators, all driven in phase by a radiofrequency (RF) signal generator. The resulting signal is then subsequently amplified with a semiconductor optical amplifier (SOA), sent to a programmable waveshaper, and amplified again with a ytterbium-doped fiber amplifier (YDFA) before its injection into the DOPO with a lensed fiber. The programmable waveshaper is used for dispersion compensation and allows the formation of pulses from 1 to 10\,ps. The RF signal generator frequency is initially chosen to match the cavity FSR (i.e., $\Delta t=0\,$fs/roundtrip). For this purpose, we scan the pump center wavelength to reveal the DOPO resonances and choose the RF signal generator frequency that maximizes their amplitudes. This maximization is obtained when the pump repetition rate coincides with the resonator FSR ($\Delta t=0\,$fs/roundtrip)\,\cite{roy_visible--mid-ir_2023}. From this synchronous regime, the RF signal generator frequency, or equivalently, the pump repetition rate, can be modified to access the topological soliton regime. The pump center frequency (1040\,nm to 1065\,nm) is chosen to work at the degeneracy and is then slightly tuned to match the closest OPO resonance ($\Delta\approx 0$). The quasi-pulsed pump regime is achieved by modulating the current of the SOA\,\cite{ledezma_octave-spanning_2023}. Specifically, the SOA was turned on during 100\,ns every 100\,$\mu$s. The DOPO output signal is sent to the degenerate optical parametric amplifier (DOPA) chip to measure its temporal profile.

\subparagraph*{\hskip-10pt Temporal measurements}
The DOPA chip consists of a 6\,mm-long lithium niobate nanophotonic poled waveguide\,\cite{zacharias_energy-efficient_2025}, phase-matched for degenerate optical parametric amplification (or second-harmonic generation at 2090 nm) at 1045\,nm\,\cite{jankowski_ultrabroadband_2020,ledezma_intense_2022}.
It is dispersion-engineered to have a zero-GVM between the signal (2090\,nm) and the second-harmonic, as well as a GVD close to zero near these two wavelengths. 
The advantages of using a nanophotonic DOPA are three-fold. First, the exponential amplification arising from optical parametric amplification/difference frequency generation allows the signal to be as low as a few microwatts. Second, its gain bandwidth exceeds hundreds of nanometers. Finally, it reduces the required gate signal power to tens of milliwatts. 
The gating signal consists of 100-fs-long pulses centered at 1045\,nm, generated by a commercial mode-locked laser. Its repetition frequency is slightly tunable ($250\pm1\,$MHz). Since the DOPO FSR is around 8.158\,GHz, we lock the 65th harmonic of its repetition rate to the pump frequency with a $\Delta f_\mathrm{rep} = 5\,$Hz offset using a proportional-integral-derivative (PID) controller. This dual-comb configuration introduces the variable delay required for cross-correlation (X-CORR) and is similar to the delay arising from a scanning stage in a conventional cross-correlation experiment. 
To measure the temporal profile around the signal frequency ($\omega_1$), we first filter the residual pump (EO comb) and send the DOPO signal and the gate pulses to the DOPA chip using a dichroic mirror (DM). A tiny fraction of the DOPO signal is also sent to an optical spectrum analyzer (OSA) with a beamsplitter (BS). After the DOPA chip, we remove the gate pulses using a filter and measure the light near the signal frequency ($\omega_1$) with a slow photodetector. We finally reconstruct the signal intensity envelope with an overall temporal resolution of 5\,fs. The DOPO pump temporal profile is measured with a standard table-top second harmonic generation cross-correlation setup\,\cite{trebino_frequency-resolved_2000}. To this end, the DOPO output and the gate pulses are redirected using magnetic mirrors (MM).

\section*{Author contributions}
N.E. and A.M. conceived the project. N.E. and R.M.G. performed the experiments. T. Z. and R. R. assisted with the temporal measurements using the DOPA chip. N.E. simulated the mean-field model and analyzed the results. L.L. and R.S. designed, fabricated, and characterized the DOPO and the DOPA chips used in the experiments. B. K. G. assisted with DOPO chip design. P. P.-R. performed the theoretical analyses. N.E. and A. M. wrote the manuscript with inputs from all authors. All authors discussed the results and contributed to the final manuscript. N.E. and R.M.G. equally contributed to the project. A.M. supervised the project. 

\section*{Acknowledgments}
The authors thank F. Leo for fruitful discussions. The device nanofabrication was performed at the Kavli Nanoscience Institute (KNI) at Caltech. The authors gratefully acknowledge support from DARPA award D23AP00158, ARO grant no. W911NF-23-1-0048, NSF grant no. 2408297, 1918549, AFOSR award FA9550-23-1-0755, the Center for Sensing to Intelligence at Caltech, the Alfred P. Sloan Foundation, and NASA/JPL. N.E. acknowledges support from the Belgian American Educational Foundation (B.A.E.F.) and the European Union’s Horizon Europe research and innovation programme under the Marie Skłodowska-Curie Grant Agreement No. 101103780.

\section*{Data Availability}
The data that support the findings of this study are available from the corresponding author upon reasonable request.

\section*{Competing Financial Interests statement}
R.M.G., B.K.G, L.L., and A.M. are inventors on a U.S. patent application US18/543,950. L.L. and A.M. are inventors on a U.S. patent 11,226,538. R.S., L.L., and A.M. are involved in developing photonic integrated nonlinear circuits at PINC Technologies Inc. R.S., L.L., and A.M. have an equity interest in PINC Technologies Inc. The remaining authors declare no competing interests. \newpage

\bibliography{Ref} 
\bibliographystyle{naturemag}

%\foreach \x in {1,...,8}
%{%
%\clearpage
%\includepdf[pages={\x}]{SM} 
%}

\end{document}